\documentclass[10pt, a4paper]{article}
\usepackage[subpreambles=true]{standalone}
\usepackage[utf8]{inputenc}
\usepackage{import}
\usepackage{hyperref}
\usepackage{enumitem}
\usepackage{setspace}
\usepackage{geometry}
\usepackage{amsmath}
\usepackage{graphicx}
\usepackage{braket}
\hypersetup{colorlinks=true, citecolor=blue, urlcolor=blue, linkcolor=blue}
\usepackage[square,numbers,sort]{natbib}
\usepackage{graphicx}
\usepackage{caption}
\usepackage{subcaption}
\usepackage{url}
\begin{document}
\title{
The QQUIC Transport Protocol: Quantum assisted UDP Internet Connections}
\author{Peng Yan, and Nengkun Yu\\
CQSI, FEIT, University of Technology Sydney, Australia
}

\maketitle
\begin{abstract}
Quantum key distribution, initialized in 1984, is a commercialized secure communication method which enables two parties to produce shared random secret key by the nature of quantum mechanics. We propose QQUIC (Quantum assisted Quick UDP Internet Connections) transport protocol, which modifies the famous QUIC transport protocol by employing the quantum key distribution instead of the original classical algorithms in the key exchanging stage. Thanks to the provable security of quantum key distribution, the security of QQUIC key does not depend on computational assumptions. Maybe surprisingly, QQUIC can reduce the network latency in some circumstance even comparing with QUIC. To achieve this, the attached quantum connections are used as the dedicated lines for key generation.
\end{abstract}

\section{Introduction}
\par The problem of generating a secret key between two remote parties affected anyone wishing to use encryption in modern communication. Traditionally, symmetric encryption suffered one enormous shortcoming – it was necessary for either the sender or the recipient to create a key and then send it to the other party. While the key was in transit, it could be stolen or copied by a third party who would then be able to destroy the encryption.

 \par In the 1970s, the Diffie–Hellman key agreement method \cite{10.1109/TIT.1976.1055638}, a method of distributing keys without actually sending the keys themselves was developed. It provides the basis for a variety of authenticated protocols and is used to provide forward secrecy in the famous Transport Layer Security (TLS)'s ephemeral modes.

\par In modern internet engineering, much effort has been focused on establishing a faster and secure connection by employing TLS protocols. Deployed in 2012 by Google \cite{10.1145/3098822.3098842}, QUIC (pronounced 'quick') is a general-purpose transport layer network protocol based on UDP to improves the performance of connection-oriented web applications based on TCP with greatly reduced latency. The TLS 1.2 was used in QUIC to provides transport security to a connection. QUIC is used by more than half of all connections from the Chrome web browser to Google's servers.
\par Both the RSA and Elliptic Curve Diffie-Hellman asymmetric algorithms which set up the TLS exchange would be vulnerable to quantum computers big enough to run Shor's algorithm. For instance, the Diffie–Hellman key agreement itself is a non-authenticated key-agreement protocol. In other words, the security of the Diffie–Hellman key agreement method, as well as RSA, is threatened by Shor's polynomial-time algorithm for discrete logarithms problem \cite{Shor:1997:PAP:264393.264406} which can be performed on the emergent quantum computer. Elliptic-curve Diffie–Hellman (ECDH) Digital Signature Algorithm (DSA), Elliptic Curve Digital Signature Algorithm (ECDSA). Shor’s algorithm can make modern asymmetric cryptography collapse since it is based on large prime integer factorization or discrete logarithm problem. Therefore, it is not considered as a Quantum-Safe or Post-Quantum cryptographic algorithm.

\par Thanks to quantum mechanics, Bennett and Brassard introduced quantum key distribution (QKD) \cite{BEN84,Bennett1992} in 1984, which enables two remote end nodes to establish provably-secure random keys. To implement quantum key distribution, it is sufficient for the quantum processors to be capable of preparing and measuring only a single qubit at a time. Many research teams have succeeded in building and operating quantum cryptographic devices since the last century. In \cite{Elliott:2003:QCP:863955.863982}, the world’s first network, the DARPA Quantum Network, was built, which delivered end to end network security via high-speed QKD by BBN, Harvard, and Boston University. Building the 1,200-mile quantum communication landline between Beijing and Shanghai in 2016 and the quantum communication satellite (known as Micius) in 2017, China has the world’s first space-ground quantum network \cite{YCL+17}. Samsung and SK Telecom have just announced the Galaxy A Quantum smartphone, packing a quantum random number generation (QRNG) chipset for improved security, where a QRNG chip is capable of generating truly random and unpredictable numbers. The QRNG chip has an onboard CMOS image sensor to detect photons, with this being the basis for the random number generation used for encryption keys.

\par To the best of our knowledge, QKD has not been employed as a component in the design and implementation of modern internet protocols. This is very surprising regarding the fundamental role of key distribution playing in communication security. In this paper, we propose QQUIC transport protocol by carefully adding QKD into the famous QUIC protocol, and using the shared random secret key to encrypt and decrypt messages. We develop an encrypted transport protocol to improve the security and transport performance for HTTPS traffic based on the new TLS 1.3 and quantum key distribution BB84. The basic motivation is that quantum key distribution(QKD) can generate a completely random secret shared key known only to the communication sides, which can be used to improve the security dramatically and optimize the procedure of key exchange in the conventional transport protocol. The QKD is mainly used to replace the key exchange to create the shared secret in the initial handshake, and the security of application data after the connection establishment is still undertaken by the TLS. Interestingly, QQUIC can reduce the network latency in some circumstance even comparing with QUIC.

\section{Preliminary}
\subsection{Quick UDP Internet Connection}
\par Transport Layer Security (TLS) is designed to achieve the goal of a safe connection between a client and a server by providing privacy, integrity and authentication for the transmitted data over the network. The security of data can be guaranteed via the symmetric cryptography with a known shared key, and the public-key cryptography can be used to provide the digital signature and authenticate the identity of the communicating parties, and the message authentication code (MAC) generated for every message can be used to detect message tampering and forgery.
\par Designed in 2012 by Google, QUIC has made great improvements in some aspects such as stream multiplexing, flow/congestion control, loss recovery and connection migration. The most splendid advantage of QUIC is that it has much lower connection establishment latency compared with the combination of TCP and TLS. QUIC employed the TLS handshake to drive the keys used for data encryption, and also relies on TLS for authentication, parameters negotiation and state change information. Here we mainly focus on the connection establishment process and briefly restate the full handshake of TLS used in current QUIC, as shown in Figure 1.

\begin{figure}[h]
    \centering
    \includegraphics[width=18em, keepaspectratio]{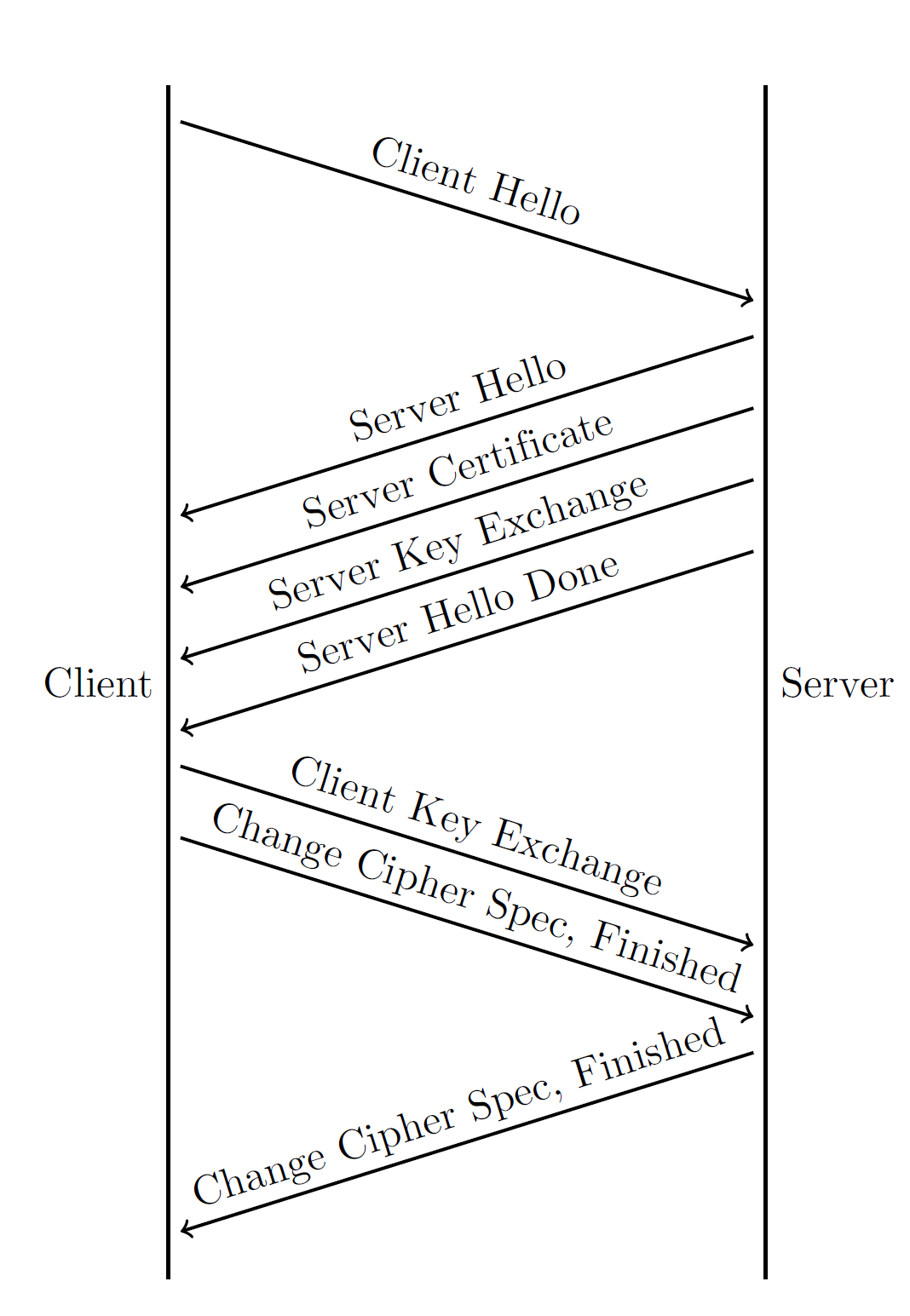}
    \caption{TLS 1.2 full handshake}
\end{figure}

\par The client sends a ClientHello(CHLO) package to initiate a new handshake and mainly contains information, a) a random client number $R_c$ for generating the master secret, b) supported TLS versions and cipher suits, c) extensions for session resumption, signature and compression algorithms. The server responds to the client with a ServerHello(SHLO) package at once, which also contains a random server number $R_s$, the choice of TLS version and cipher suits and response for extensions. Then the signed certificate is sent by the server for proving its identity to the client, which contains the server's long-term public key. Besides, a Server Key Exchange message is sent if the public key included in the server's certificate is not sufficient for the client to finish the key exchange. Finally, a Server Hello Done message is sent to indicate that the server is done and awaiting for the client's response.
\par Once the client receives the ServerHello message, it first verifies the server's certificate and caches some information needed for a reconnection. Optionally, the client may need to send its certificate if required by the server. Similarly, a Key Exchange message encrypted by the server's public key is also sent by the client. Now the client is able to infer the master secret based on the key exchange messages and the random number $R_c$ and $R_s$. A Change Cipher Spec message will be sent to notify the server that any data sent by the client from now on will be encrypted using the symmetric key derived from the master secret, followed by a Finished message which contains a hash of all handshake messages sent previously. Once the server uses its private key to decrypt the client Key Exchange message, the server can also calculate out the same master secrets just like the client. Now the server can use the symmetric key also derived from the master secret to decrypt the client handshake Finished message and check the validity of all previous handshake messages. After this, the server will also send the Change Cipher Spec and the Finished message just like the client for the same goal. Once the client successfully verifies the server Finished message, a full TLS Handshake is completed and application data can be transmitted securely. Therefore, TLS 1.2 takes 2-RTT to create a new secure connection.

\subsection{Quantum Key Distribution}
\par BB84 \cite{BEN84} is a quantum key distribution scheme developed by Charles Bennett and Gilles Brassard in 1984. The protocol is provably secure \cite{PhysRevLett.85.441}, relying on the fundamental aspect of quantum mechanics that information gain is only possible at the expense of disturbing the signal \cite{DIEKS1982271,Wootters1982Single}.The workflow of BB84 is shown in Figure 2.

\begin{figure}
\centering
    \includegraphics[width=10cm]{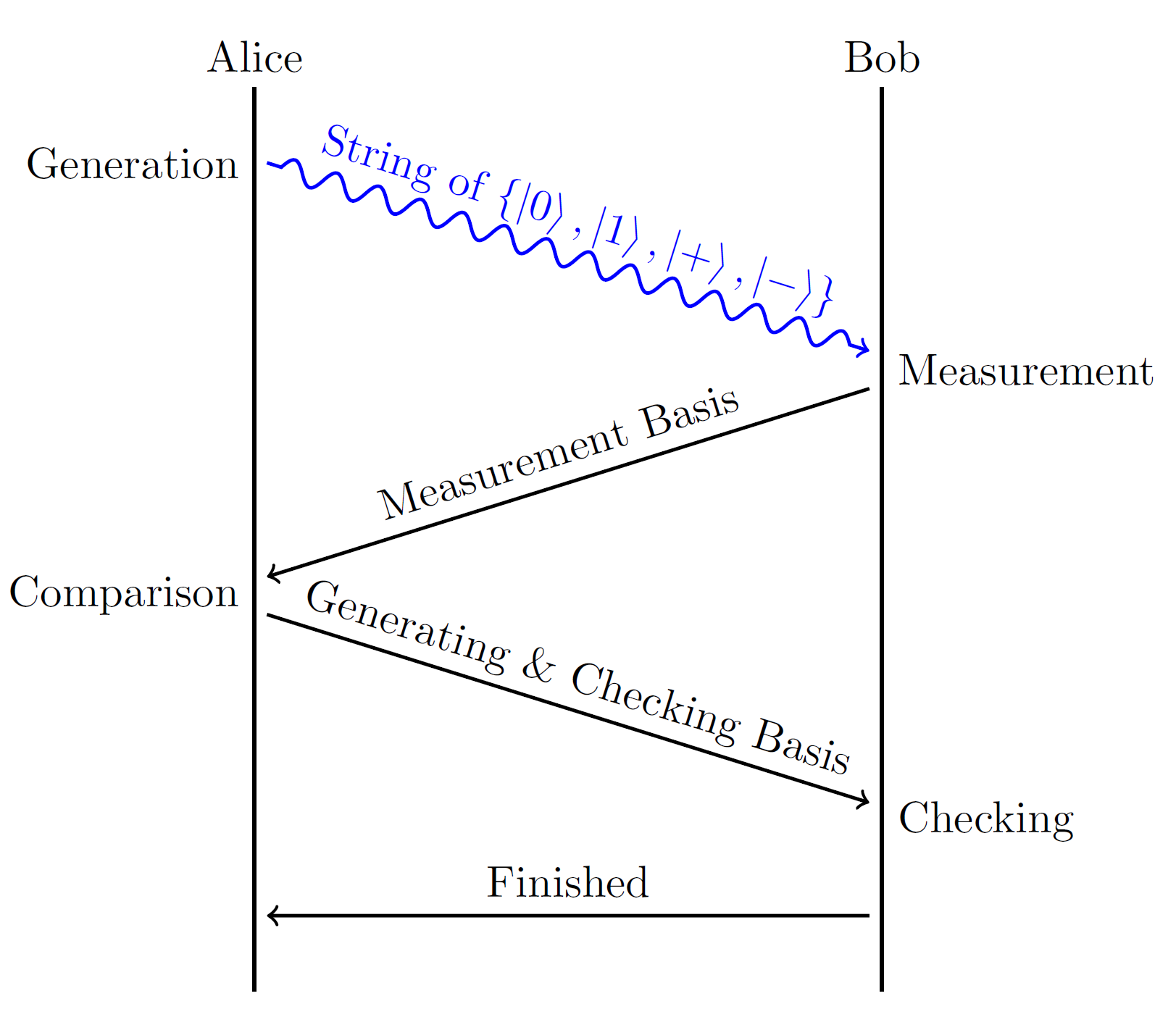}
    \caption{BB84 workflow. A string of polarized photons randomly in $\{\ket{0}, \ket{1}, \ket{+}, \ket{-}\}$ sates are sent by Alice to let Bob give random measurements in sequence. After the measurement on the photons and basis communication, both sides will get a shared secret.}
\end{figure}

\begin{enumerate}
\item Alice randomly generates two strings, $a=a_0a_1\cdots a_n$, and $b=b_1b_2\cdots b_n$. $a$ is called the basis string. If $a_i=0$, she chooses a qubit in $\{\ket{0}, \ket{1}\}$ according to $b_i$; otherwise, she chooses a qubit in $\{\ket{+}, \ket{-}\}$\footnote{$\ket{+}=\frac{\ket{0}+\ket{1}}{\sqrt{2}}$ and $\ket{-}=\frac{\ket{0}-\ket{1}}{\sqrt{2}}$.} according to $b_i$. These qubits are sent by Alice to Bob in strict sequence.
\item After receiving these quantum states, Bob randomly measures each qubit in basis $\{\ket{0}, \ket{1}\}$ or in basis $\{\ket{+}, \ket{-}\}$, and records his choice of basis string as $a'$. After he gets the measurement outcome $b'$, $a'$ was sent to Alice.
\item Alice discards the bits of $b$ in the locations where $a$ and $a'$ do not match. Alice announces $a$ first, then she randomly takes the half of the rest bits of $b$ to be the checking bits $b_c$ and announces the selection.
\item Bob discards the bits of $b'$ in the locations where $a$ and $a'$ do not match after knowing $a$. Then he checks if the states of selected qubits in checking bits $b_c$ before and after the measurement are the same. If less than an acceptable threshold disagree, the quantum channel is reliable and the remaining bits can be used to obtain the shared key bits.
\end{enumerate}

\section{QQUIC Transport Protocol}
\par In the last section, we have briefly explained how the BB84 works to generate a shared secret between two sides. If we want to combine these properties to the conventional transport protocols, it requires a quantum channel to generate, transport and measure the polarized photons in a predefined way, and also needs two peers to communicate their basis choice and measurements through the classical channel. It may look much more complicated compared with the traditional key exchange process, but the quantum channel and classical channel can work parallelly to finish the handshake, and QKD can also continuously work to provide keys for key updating after the connection establishment.
\par We introduce a new secure transport protocol called QQUIC which based on the QUIC with QKD replacing the key exchange stage. The encryption and transmission of the application data after connection establishment just works similarly as QUIC, except for the initial full handshake for a new connection and key updating. There are two channels during the connection, one is the classical channel just as the traditional transport protocols, another quantum channel is used for all necessary operations on the photons. The network structure of the protocol works as Figure 3.

\begin{figure}[h]
\centering
    \includegraphics[width=30em, keepaspectratio]{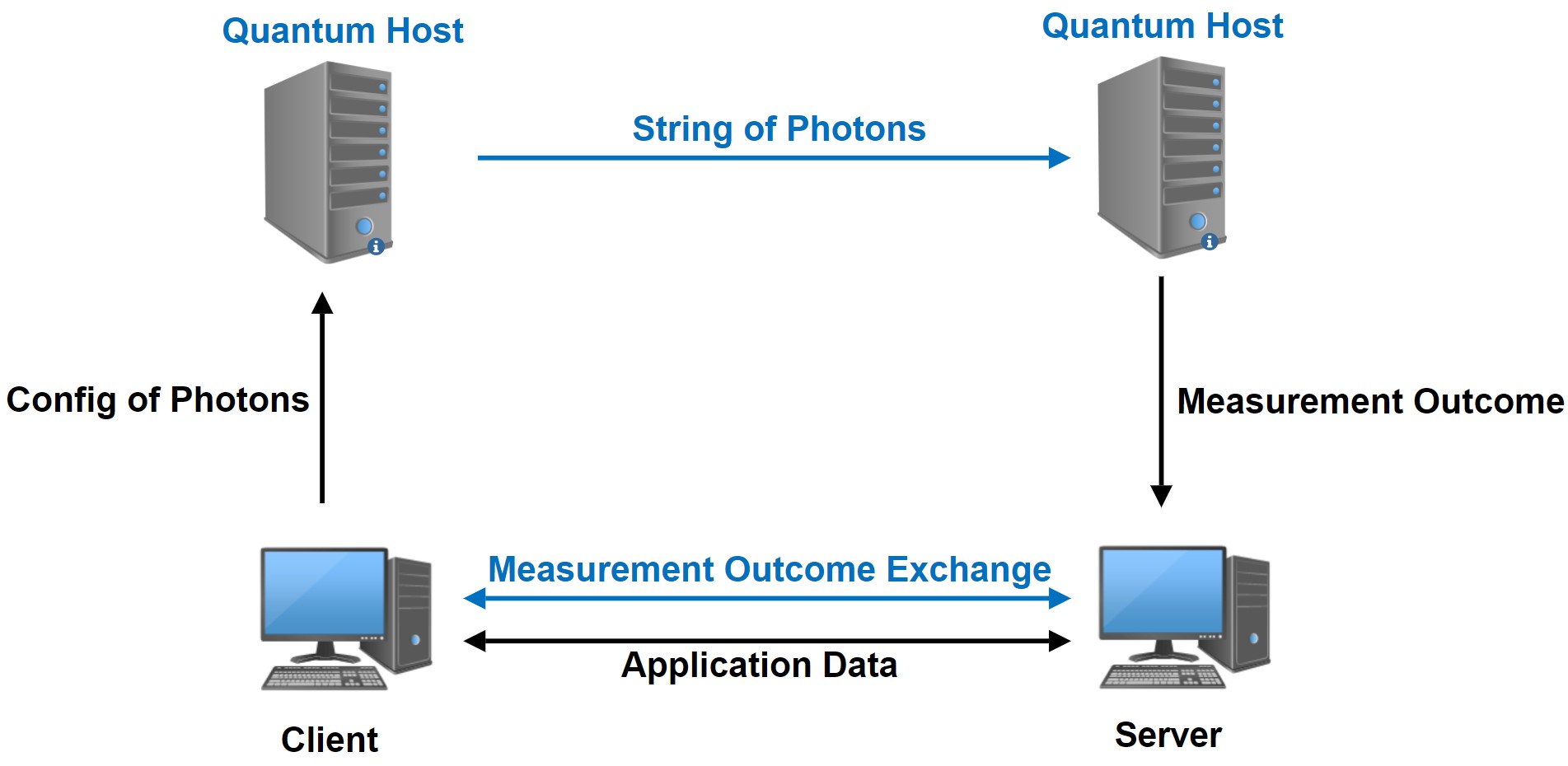}
    \caption{Network Structure of the QQUIC}
\end{figure}

\par Each side has a quantum host and a classical host. The quantum hosts and their authenticated quantum channel are responsible for generating the shared secret to replace the classical key exchange. For the initial connection handshake and then periodical key updating, the client needs to activate its quantum host to generate and send strings of special polarized photons, and the task for the generation of true randoms $a$ and $b$ which determine the photon states can be also authorized to the quantum host. Meanwhile, the client informs the configuration of the photons to the server through the classical channel. After the server knows the configuration of these photons, it's quantum host can choose a basis string $a'$ and use it to get the measurement outcome $b'$ of received photons. After the measurement, two sides communicate their basis choice $a$ and $a'$ and discards the useless bits in $b$ and $b'$. After verifying the validity of the quantum channel, the secure shared secret can be created and used for key derivation, where it takes 1-RTT to make the shared secret negotiation in the classical channel. Here we give a detailed description of the workflow of the full handshake in QQUIC, shown in Figure 4.

\begin{figure}[h]
\centering
    \includegraphics[width=18em, keepaspectratio]{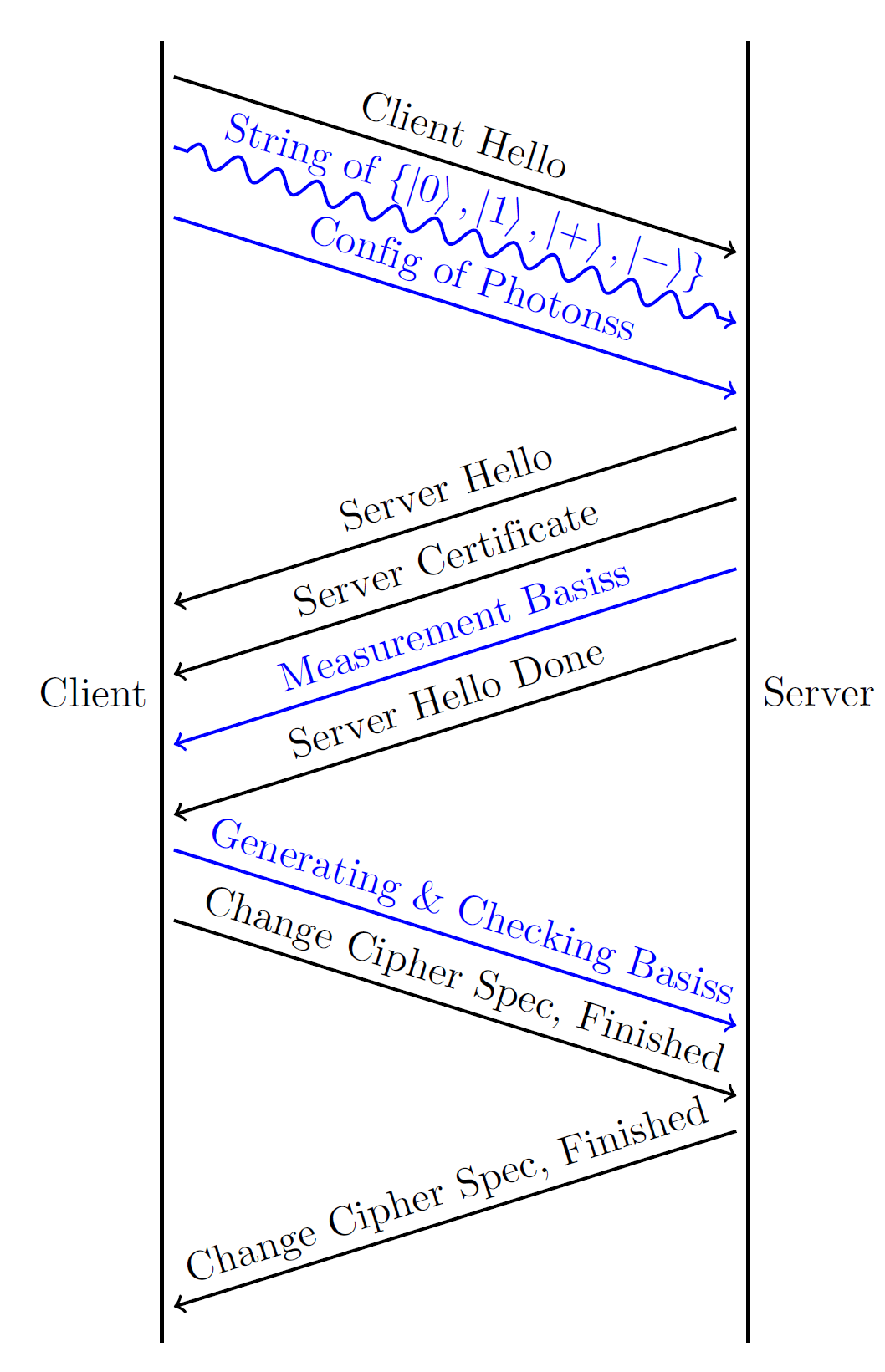}
    \caption{QQUIC full handshake}
\end{figure}

\par Initially, the client has no information about the server, so it needs to initiate the connection via a ClientHello(CHLO) message. The CHLO message contains the lists for supported TLS versions, cipher suites, signature algorithms, and extension for session resumption, which is similar to the CHLO message in TLS 1.3. However, the CHLO package has no information related to key exchange such as a random client number or key exchange extension. Instead, the client requires its quantum host to generate two private binary randoms $a$ and $b$ to determine how to generate a string of polarized photons and send them through the quantum channel at the same time, as explained in BB84. Thus the CHLO message usually needs to contain some extra configure information about the photons, such as the set of four possible basis used for photons generations, the amount and sending speed of photons. The transmission of photons in the quantum channel works parallelly with the CHLO message in the classical channel.
\par The server gives a ServerHello(SHLO) message as a response to the CHLO package. The SHLO message contains: a) the server configuration information, b) the choice of cipher suites and signature algorithms, c) a certificate and certificate verify message. In the meantime, the server's quantum host needs to receive these polarized photons in the same order, and measure them in sequence with a basis string $a'$ which is randomly selected from two possible bases provided by client. After this, the server keeps the measurement results $b'$ privately and sends the string $a'$ (measurement basis) to the client through the classical channel.
\par Once the client receives the SHLO message from the server, it starts to verify the identity of the server via certificate and signature. After this, the client has to wait for the arrival of handshake message about the server's choice of measurement basis, and discard the bits in $b$ where two sides choose different basis. Besides, the client needs to randomly take half of the remaining bits in $b'$ to create a random $b_c$, and the rest of bits in $b'$ can be used to obtain shared secret. Now the client needs to send $a$ (generating basis) to inform the server the basis string for generating the photons, and $b_c$ (checking basis) for server to decide which photons are used for checking. Finally, the client should use the derived key to encrypt the Change Cipher Spec message and the Finished message which contains a hash of all the handshake information up to this point to avoid handshake message tampering.
\par After receiving handshake messages from the client, the server should first check if the states after measurement on the photons related to the bits in $b_c$ are the same as the corresponding initial states send by client which can be inferred from $a$ and $b_c$. If the difference rate is above a safe threshold, it means there are too much noise or exists the presence of an eavesdropper in the quantum channel, thus the handshake has failed and another try is a must. Then the server can infer the same shared secret based on the $a$ and $b_c$ from client and its own measurement results $b'$, and use it to decode the Finished message from the client to check the rightness of all the previous information. In the end, the server also sends the Change Cipher Spec and Finished message as the client. After the client checks this Finished message, a safe connection has been established.
\par As a comparison, the establishment of a safe connection in this protocol requires 2-RTT, which is 1-RTT slower than TLS 1.3. However, if the success rate of the validity checking for the quantum channel is high enough, the client can also send some unimportant encrypted application data (which is the common case) immediately after the client's Finished message in the first round trip, and the server can also directly give its response to the request after its Finished message. This "quick start" strategy can be used in most HTTP request which usually begins with an unimportant request such as permission for log-in. As for the session assumption, PSK mode and 0-RTT reconnection used in QUIC can be also deployed here. Again, because the quantum channel needs 2-RTT to generate the shared secret safely, the application data in the first two round trips during the handshake are still encrypted by the previous old key.
\par It has been confirmed that a connection encrypted by one session key for a long time is not safe enough even using authenticated encryption with associated data(AEAD). In order to guarantee security, TLS 1.3 proposes to use the KeyUpdate message to update session keys in an existed connection, which takes 1-RTT to notify each side and derive the next shared traffic secret based on current traffic secret and KeyUpdate message, as shown in Figure 5(a).

\begin{figure*}[h]
    \centering
    \begin{subfigure}[b]{0.4\textwidth}
        \centering
        \includegraphics[width=\textwidth, keepaspectratio]{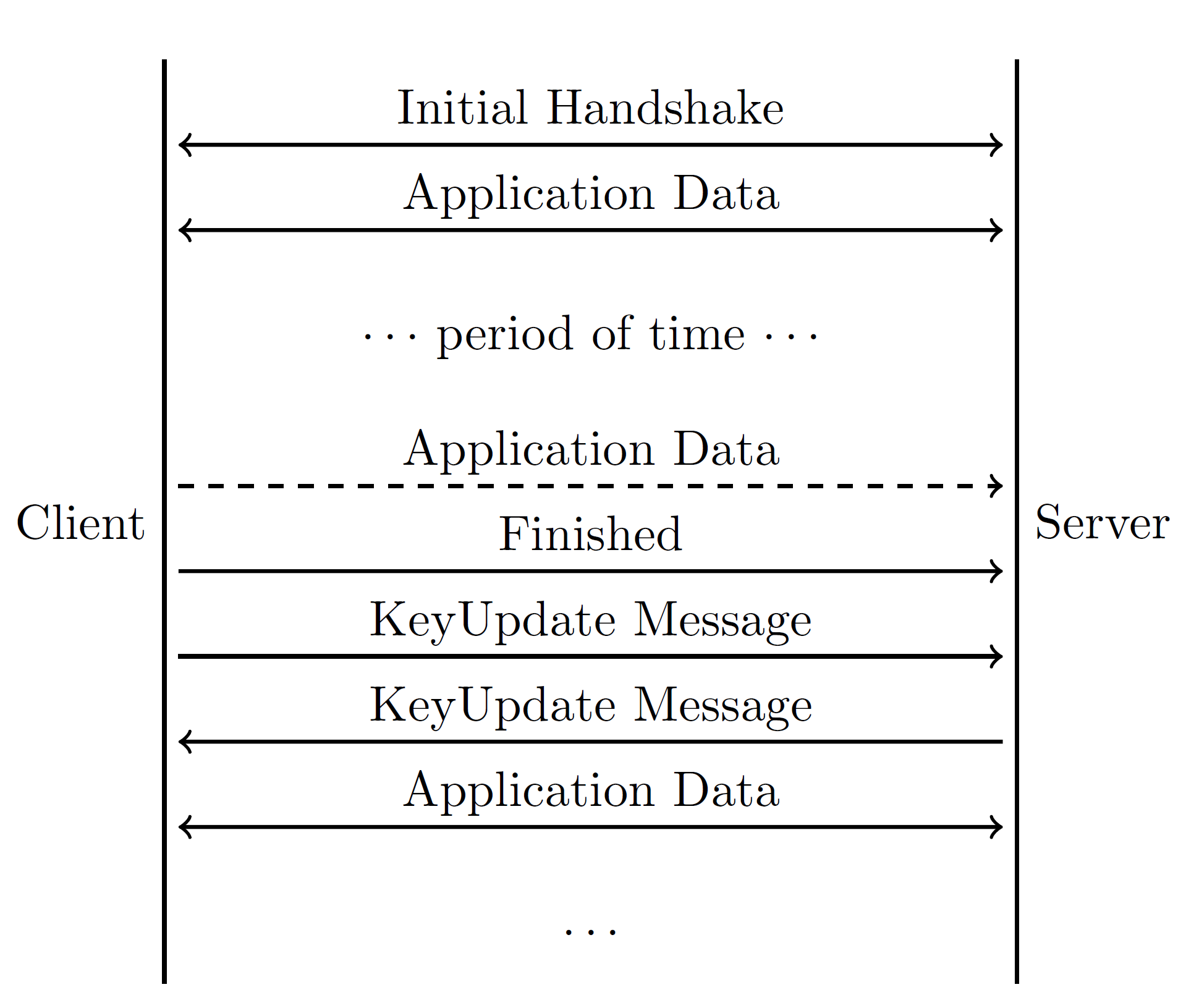}
        \vspace{3.3pt}
        \caption{KeyUpdate in TLS 1.3}
    \end{subfigure} %
    \begin{subfigure}[b]{0.4\textwidth}
        \centering
        \includegraphics[width=\textwidth, keepaspectratio]{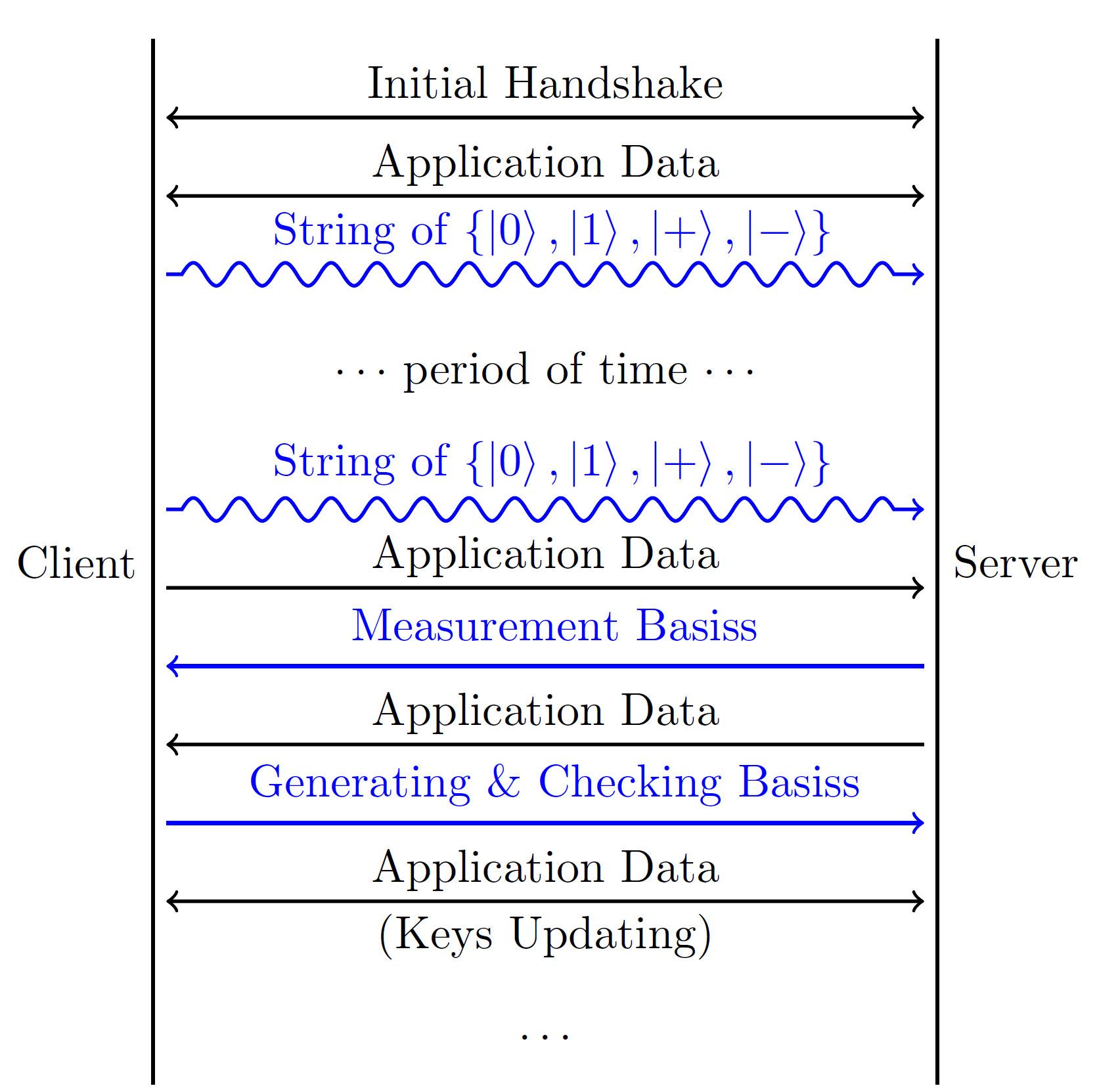}
        \caption{KeyUpdate in QQUIC}
    \end{subfigure}
    \caption{KeyUpdate in a long-term connection}
\end{figure*}

\par Besides the much higher security level of the master secret from QKD, the most obvious advantage of this new protocol is that the key generation and application data transmission can work parallelly. As shown in Figure 5(b), after the connection establishment, the client can keep sending strings of photons continually and each string can generate a master secret. The server measures a group of several photon strings once and negotiates these generation and measurement bases in one round trip, which can effectively save the travel time. This would keep each side always have several shared secrets stored in advance for the future key update. If both sides have made an agreement about the number of ordered packages for the lifetime of one session key, then two sides can automatically switch prepared session keys in sequence without notice when the right time comes. The package number increased with time and stream offset in QUIC are used to reorder received packages and choose the right session key for different package chunks. This protocol will have huge advantages in long-term connections such as video conference and big data transmission which requires frequent key updates. And due to the higher security of QKD, the results from QKD can be used as master secret directly, and there is no need to put more random materials to generate the final session keys compared with the HKDF used in TLS 1.3. The same goes for the frequent key updates in the long-term connection, where the session key derivations of different chunks are totally independent.

\section{Discussion and Conclusion}
\par In this paper, we propose a new protocol named QQUIC which combines the advantages of QUIC and quantum key distribution, to achieve more secure and effective key generations for the initial handshake and key updating. This protocol uses QKD to replace the classical key exchange procedure and performs a 2-RTT full handshake for a new connection, and even 1-RTT time cost with "quick start" technique. The concurrent work of the quantum and classical channels makes the key generations much more efficient and resource saving, which has huge advantages in the long connection applications such as video meeting and database share. The detailed implement and simulation will be carried out in the near future to check out the validity of this new protocol.
\par For an implementation of QKD with relatively high key rate, the portion of asymmetric cryptography in the network protocol may be significantly reduced. This would provide a considerable computational speedup due to a well known fact that public-key cryptography, or asymmetric cryptography, is much slower than symmetric cryptography.
\par NY thanks Gushu Li for helpful discussions at the beginning of this project. NY is grateful to Zhengfeng Ji for sharing his insight on this topic. We are thankful for Mingsheng Ying's constant support and encouragement.
This work is supported by DE180100156.

\bibliographystyle{plain}
\bibliography{ref}
\end{document}